# Specific heat in the superconducting and normal state (2-300 K, 0-16 Teslas), and magnetic susceptibility of the 38-K superconductor MgB$_2$: evidence for a multicomponent gap.

Yuxing WANG, Tomasz PLACKOWSKI[†] and Alain JUNOD

Département de Physique de la Matière Condensée, Université de Genève,
CH-1211 Genève 4 (Switzerland)

(March 7, 2001)

The specific heat C of a sintered polycrystalline sample of MgB$_2$ with a bulk superconducting transition temperature $T_c$= 36.7 K is measured as a function of the temperature (2-300 K) and magnetic field (0-16 T), together with magnetic properties (normal-state susceptibility, superconducting state magnetization, etc.). The Sommerfeld constant $\gamma = 0.89 \pm 0.05$ mJ/K$^2$/gat (2.7 mJ/K$^2$/mol) is determined in the normal state above $H_{c2}$. The normal- and superconducting state entropies are equal at $T_c$. Several moments of the phonon density of states are obtained from the lattice specific heat. We report bulk values for: the thermodynamic critical field, $B_c(0) = 0.26$ T; the slope of the upper critical field, $(dB_{c2}/dT)_{Tc} = 0.56$ T/K; the Ginzburg-Landau parameter, $\kappa = 38$; the coherence length, $\xi \cong 5$ nm; the lower critical field, $B_{c1} \cong 0.018$ T; the London penetration depth, $\lambda(0) \cong 180$ nm. These results characterize MgB$_2$ as a type-II superconductor. The nearly quadratic dependence of C(T) versus T at T << $T_c$, its non-linear field dependence, and the discrepancy between the electron-phonon coupling constant $\lambda_{ep}$ as determined by the renormalization of the electron density-of-states ($\lambda_{ep} \cong 0.6$) and by McMillan's equation for isotropic superconductors ($\lambda_{ep} \cong 1.1$), are inconsistent with a single isotropic gap. In addition to high phonon frequencies, anisotropy or two-band gap structure may explain why the critical temperature of this superconductor is high in spite of its low condensation energy, which does not exceed 1/16 of that of YBa$_2$Cu$_3$O$_7$ and 1/4 of that of Nb$_3$Sn.

## Introduction.

The discovery of the superconductor MgB$_2$ with a critical temperature $T_c$ = 39 K [1] has again put into question the limits of $T_c$ for superconductors in which the electron-phonon interaction provides the coupling mechanism leading to the formation of Cooper pairs. Traditional evaluations set a limit of about 30 K for this class of superconductor [2]. The origin of superconducting coupling in cuprates with critical temperatures as high as 160 K is still controversial, and may be of magnetic origin; in contrast, the recent observation of a large isotope effect in MgB$_2$ has provided strong evidence in favor of phonon-mediated coupling [3].

BCS theory and its subsequent refinements based on Eliasberg equations show that high critical temperatures in phonon-mediated superconductors are favored by high phonon frequencies and by a large density of states at the Fermi level. The Allen-Dynes ("AD") formula [4], which is an interpolation based on numerical solutions of Eliashberg equations valid over a wide range of coupling strengths and for various shapes of the phonon spectrum, expresses these relations quantitatively:

$$T_c = \frac{\langle\omega_{\ln}\rangle}{1.20}\exp\left(-\frac{1.04(1+\lambda_{ep})}{\lambda_{ep}-\mu^*-0.62\lambda_{ep}\mu^*}\right)$$

$$\lambda_{ep} \equiv 2\int_0^\infty \alpha^2 F(\omega)d\ln\omega = \frac{N(0)\langle I^2\rangle}{M\overline{\omega}_2^2}$$

$$\overline{\omega}_2 \equiv \left(\frac{2}{\lambda_{ep}}\int_0^\infty \omega^2\,\alpha^2 F(\omega)\,d\ln\omega\right)^{1/2}$$

$$\omega_{\ln} \equiv \exp\left(\frac{2}{\lambda_{ep}}\int_0^\infty \ln\omega\,\alpha^2 F(\omega)\,d\ln\omega\right)$$

where $\lambda_{ep}$ is the dimensionless electron-phonon coupling parameter, $\mu^* \cong 0.1$ is the retarded Coulomb repulsion parameter, $F(\omega)$ is the phonon density of states (PDOS) normalized to 1, $\alpha^2 F(\omega)$ is the spectral electron-phonon interaction function, $N(0)$ is the density of states at the Fermi level per spin and per atom (EDOS), $\langle I^2\rangle$ is the properly averaged electron-ion matrix element squared, and $M$ is the average atomic mass. $N(0)\langle I^2\rangle$ is the so-called Hopfield electronic parameter, whereas $M\overline{\omega}_2^2$ is an average force constant. Correction factors for very strong coupling ($\lambda_{ep} > 3$) are omitted. The essential difference with McMillan's equation [2] is the use of $\omega_{\ln}$ rather than the Debye temperature as a characteristic phonon energy.

Specific heat (C) measurements give access to several of these parameters. The specific heat jump at $T_c$ provides a measurement of the bulk critical temperature. Unlike resistivity measurements, which are sensitive to the first percolation path, and susceptibility measurements, which are sensitive to shielding by a superficial layer of superconducting material, the specific heat jump is proportional to the superconducting volume, and its width mirrors the distribution of $T_c$ within the whole volume.

The Sommerfeld constant $\gamma = \lim_{T=0} C_n/T$ is another important parameter given by the specific heat. $C_n(T)$, the normal-state curve, can be measured in a field larger than the upper critical field $H_{c2}(0)$ (if available). $\gamma$ is proportional to the EDOS renormalized by the electron-phonon interaction:

$$\gamma = \frac{2}{3}\pi^2 k_B^2 N(0)(1+\lambda_{ep})$$

where $k_B$ is the Boltzmann constant and $N(0)$ is the band structure EDOS for *one* spin direction. $N(0)$ expressed in states per eV, per atom and per spin direction, and $\gamma$ in mJ/K$^2$gat, this formula becomes $N(0)(1+\lambda_{ep}) = 0.212\gamma$.

The lattice specific heat provides information on the phonons. The initial slope of the total specific heat $\beta_3 = \lim_{T=0} d(C_n/T)/d(T^2)$ gives the initial Debye temperature $\Theta_0$

$$\beta_3 = \frac{12}{5}R\pi^4\Theta_0^{-3}$$

which depends on the velocities of sound; R being the ideal gas constant. Again, in practical units, $\beta_3 = 1944000/\Theta_0^3$ if $\beta_3$ is given in mJ/K$^4$gat and $\Theta_0$ in K. Deviations from the Debye model are common, and MgB$_2$ is no exception, so the initial Debye temperature is generally not representative of the frequencies that are most effective for pairing (i.e., typically $\hbar\omega = 2\pi k_B T_c$ [5]).

More representative averages can be defined if $F(\omega)$ is known. The spectral coupling function $\alpha^2 F(\omega)$ that enters AD's formula is not directly available without e.g. tunneling experiments, but $F(\omega)$ is sufficient if one assumes $\alpha^2 F(\omega) \propto F(\omega)$:

$$\left(\langle\omega\rangle/\langle\omega^{-1}\rangle\right)^{1/2} \approx \overline{\omega}_2$$

$$\exp\left(\langle\omega^{-1}\ln\omega\rangle/\langle\omega^{-1}\rangle\right) \approx \omega_{\ln}$$

where

$$\langle \phi(\omega) \rangle \equiv \int_0^\infty \phi(\omega) F(\omega) d\omega$$

In this approximation, $\langle \alpha^2 \rangle \cong 0.5 \lambda_{ep} / \langle \omega^{-1} \rangle$. The moments $\langle \omega^n \rangle$ can be obtained by an inversion procedure of the specific heat [6] [7] or equivalently from some particular integrals of the lattice specific heat $C_{ph}(T)$ [8] [9]. This follows from the fundamental relation between the phonon specific heat and the PDOS:

$$C_{ph}(T) = 3R \int_0^\infty E(\omega/T) F(\omega) d\omega,$$

$$E(x) \equiv x^2 e^x (e^x + 1)^{-2}$$

where $k_B E(x)$ is the Einstein specific heat of one mode with frequency $\omega$ at the temperature T, and $x \equiv \omega/T$ (we express frequencies in Kelvin, omitting the implicit factor $\hbar/k_B$). Although finding the precise PDOS from the specific heat is an ill-conditioned problem, the determination of a smoothed PDOS or of moments of the PDOS is a stable problem. As a rule of thumb, in order to probe the full phonon spectrum up to a frequency $\omega$, one needs specific heat data up to a temperature T ≈ 0.2$\omega$. The lattice specific heat must be separated from other contributions. For superconductors, the preferred method consists of measuring the normal-state specific heat in high magnetic fields and then subtracting the electronic contribution determined at T → 0. Corrections due to the difference between the constant-volume and the constant-pressure specific heat, to the temperature-dependence of γ or its renormalization are generally small.

Specific heat measurements then give information on the electron-phonon coupling strength. The ratio $T_c/\omega_{ln}$ and the AD formula determines $\lambda_{ep}$. The dimensionless specific heat jump $\Delta C/\gamma T_c$ also depends on $\lambda_{ep}$, and varies from 1.43 in the weak-coupling limit (e.g. for the superconductor Al with $T_c$ = 1.19 K) to about 3 for strong-coupling superconductors. Table II recalls characteristic parameters for Nb$_3$Sn, YBa$_2$Cu$_3$O$_7$, and lists the essential results obtained in this work for MgB$_2$.

One should keep in mind that McMillan's or AD's equations are interpolation formulas which represent the solutions of *isotropic* Eliashberg equations. An anisotropic s-wave gap which varies between $\Delta_{min}$ and $\Delta_{max}$ as a function of $\vec{k}$ would qualitatively affect the specific heat much in the same way as a distribution of critical temperatures between $T_{cmin}$ and $T_{cmax}$ ($\equiv T_c$). For a given coupling strength and condensation energy, the thermodynamic critical field would increase more slowly just below $T_c$ and more rapidly at low temperature than for the isotropic case. This would result in a smaller specific heat jump and in a distortion of the $H_c(T)$ curve. Adding anisotropy to a medium to strong-coupling superconductor *simulates* weaker coupling.

Since the electron-phonon superconductor MgB$_2$ has an unexpectedly high critical temperature, we performed extensive specific heat experiments in order to answer the questions:

- how large are the average phonon frequencies? Are they large enough to account for $T_c$ only through the prefactor $\omega_{ln}$?

- how large is the EDOS N(0)? Is it large enough to balance the phononic denominator $M\overline{\omega}_2^2$ and produce a large value of $\lambda_{ep}$?

- is MgB$_2$ a strong- or weak-coupling superconductor?

- how does the electron-phonon interaction compare with other superconductors?

- is the condensation energy $E_c$ unusually large?

- is there any indication in favour of a possible anisotropy of the gap, or two-band superconductivity?

- and, more generally, what are the values of the Ginzburg-Landau parameters $\lambda$, $\xi$, $\kappa$.

For this purpose, we measured the specific heat in the normal state down to 3 K in B = 14 and 16 T. The latter curves are undistinguishable, so that we trust that these fields are larger than $H_{c2}$. This allowed us to determine the Sommerfeld constant with <5% extrapolation error. Susceptibility measurements confirmed that the EDOS is small. In zero field, we measured the specific heat from 2 to 300 K. Using a crude inversion technique, we determined the moments of the PDOS. We applied intermediate fields 0.5, 1, 2, 3, 10, and 14 T in various temperature ranges, measured the shift of the critical temperature, and the variation of the low temperature specific heat. This provided a bulk determination of the initial slope of the upper critical field, and unveiled very unusual behavior at T << $T_c$, which differs strongly from that of an isotropic s-wave type-II superconductor. We conclude that $T_c$ is enhanced both by high phonon frequencies and by a complex gap structure in $MgB_2$.

**Samples**

Commercial powder of $MgB_2$ (Alfa Aesar) with nominal 98% purity was pressed into two pellets, sealed into quartz ampoules under 1 bar argon gas at ambient temperature, and sintered for 60 hours at 850°C. No measurable weight losses occurred. One cylindrical pellet with a diameter/height ratio of about one and mass 0.1 gram was used for magnetic measurements, and one disc-shaped pellet with a mass of 0.15 gram was used for specific heat measurements.

X-ray diffractometry showed only the lines of the $AlB_2$ structure plus one extra faint unidentified line. Iron impurities were detected by microprobe analysis at the 1-2% level, in addition to some carbon and oxygen.

**Magnetic properties**

AC susceptibility at a frequency of 8 kHz and in a field of 0.1 G r.m.s. shows the onset of superconductivity at 38 K (Fig. 1). In the 10-30 K range, a broad increase of diamagnetism accompanied with a signal in the quadrature (loss) channel indicates that weak links between grains in turn become superconducting, expelling the field from of the entire volume of the porous pellet.

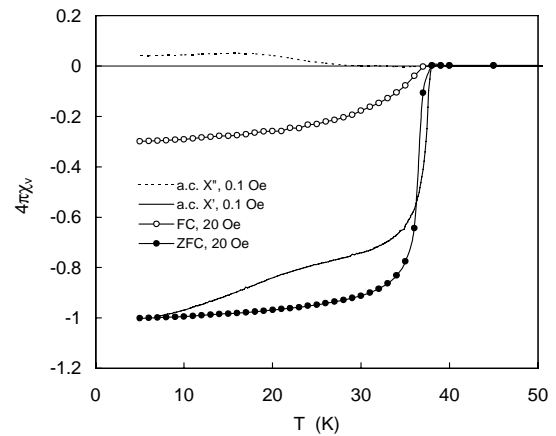

*Figure 1. Diamagnetic transition of the polycrystalline $MgB_2$ sample. Full line: AC susceptibility, in-phase $\chi'$ component. Dashed line: quadrature (loss) component $\chi''$. The signal is normalized to -1 at low temperature. Full circles: diamagnetic susceptibility in 20 G measured after cooling in zero field. Open circles: Meissner flux expulsion in 20 G measured after cooling in the same field. The DC susceptibility in emu/cm$^3$ is corrected for a demagnetization factor N = 1/3.*

The DC susceptibility measured in a SQUID magnetometer in a field of 20 Oe (zero field cooling conditions) shows full diamagnetism at low temperature ($\chi = -1/4\pi$), taking into account the demagnetization factor $N = 1/3$ of the short cylinder (Fig. 1). The normalization volume for this experiment is the measured mass divided by the X-ray density, 2.625 g/cm$^3$. 60% of the transition occurs over a 1 K interval, with an onset between 37 and 38 K. The Meissner curve (field cooling conditions, 20 Oe) shows an onset at 37 K, and 30% diamagnetism at low temperature. The Meissner fraction is reduced by irreversibility. The same powders sintered by hot-pressing at 950-1150 °C (not shown here) exhibit sharp transitions with ≈1 K full width, but surprisingly the Meissner curve becomes *paramagnetic* [10]. This is likely to be related to the ferromagnetism of Fe impurities, together with a large magnetic irreversibility. A signal corresponding to a concentration of 730 ppm ferromagnetic Fe clusters is observed above $T_c$ (Fig. 4, inset).

The large critical current density and irreversibility of the sintered samples is supported by the magnetization measurements (Fig. 2). The hysteresis at 2.5 kOe and 5 K corresponds to a critical current density of $10^4$ A/cm$^2$ in Bean's critical state model [11]. The reversible region lies above 30 kOe at T = 5 K, and above 40 kOe at T = 15 K (Fig. 2, upper inset). This precludes the use of magnetization curves for thermodynamic analyses. Only an upper limit for the lower critical field $H_{c1}$ can be obtained from these curves, by noting that the first deviation from Meissner diamagnetism occurs between 300 and 500 Oe (Fig. 2, lower inset).

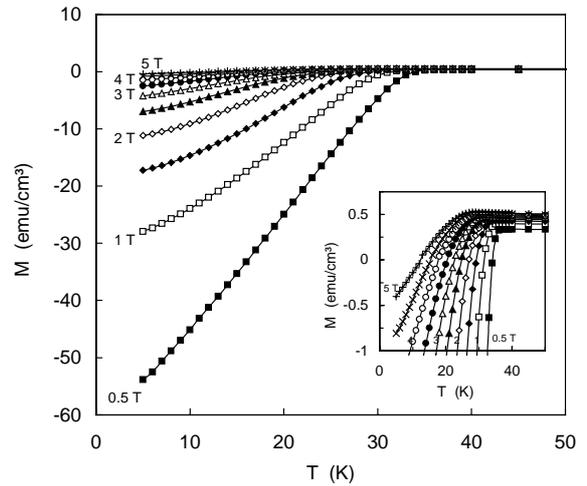

*Figure 3. Main frame: magnetization measured at constant field with increasing temperature after zero field cooling. Inset: expanded view showing the non-linear behavior as $T_{c2}(H)$ is approached.*

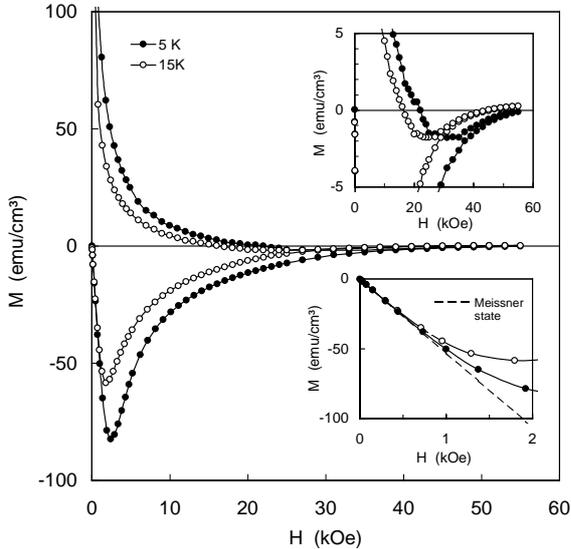

*Figure 2. Main frame: virgin magnetization curves at 5 and 15 K. Upper inset: expanded view showing the onset of the reversible regime. Lower inset: expanded view at low field showing a deviation from full diamagnetism starting between 300 and 500 Oe.*

Large differences have been found in the magnetization versus temperature at constant fields between sintered samples (Fig. 3) and hot-pressed samples. Sintered samples have typically 10 times smaller magnetization than hot-pressed ones. The prescription used to determine the upper critical field $H_{c2}(T)$, or rather $T_{c2}(H)$, in cuprate superconductors [12] is not reliable here, as it consists of extrapolating the linear part of the M(T) curve to find its intercept with the normal-state magnetization, whereas here there is a large irreversibility and non-linearity of M(T) (Fig. 3, inset). If nevertheless attempted, such a construction yields a slope $-\mu_0(dH_c/dT)_{Tc} \approx 0.2$ T/K between 0.5 and 2.5 T. This corresponds approximately to the shift of the peak of the specific heat (Fig. 8), i.e. to the lower bound of a distribution of $T_{c2}$ in B > 0.

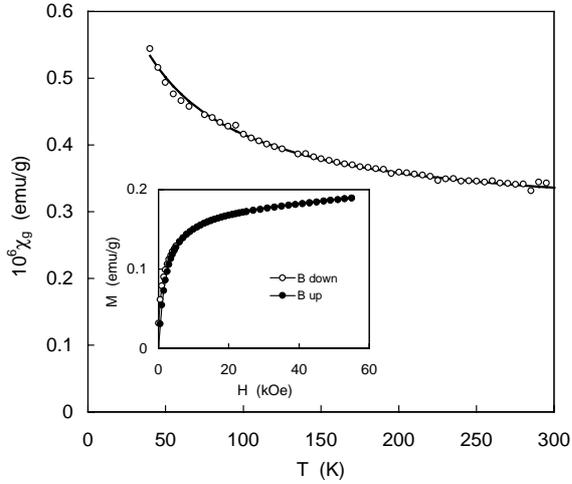

*Figure 4.* Main frame: normal-state susceptibility of $MgB_2$ versus T. Full line: fit with a Pauli susceptibility and a Curie-Weiss term. Inset: magnetization cycle at T = 50 K showing the saturation of Fe impurity moments.

Figure 4 shows the normal-state susceptibility between 40 and 300 K. The susceptibility is obtained from the high field slope:

$$\chi = \frac{M(H_2) - M(H_1)}{H_2 - H_1}$$

with $\mu_0H_2 = 5.5$ T and $\mu_0H_1 = 3$ T. This procedure approximately cancels the contribution of saturated moments of Fe clusters (Fig. 4, inset). The susceptibility can be described by $\chi = \chi_0 + C_c/(T - \Theta_c)$, where $\chi_0 = 4.4\times10^{-6}$ emu/gat, $\Theta_c = -26$ K and $C_c = 2.5\times10^{-4}$ emu·K/gat (full line in Fig. 4). Subtracting $\chi_c = -1\times10^{-6}$ emu/gat for the diamagnetic core susceptibility of $Mg^{2+}$ ions [13], one finds $\chi_0 = 5.4\times10^{-6}$ emu/gat for the Pauli susceptibility of conduction electrons. This corresponds to an EDOS N(0) = 0.083 states/eV·atom·spin, about 30% smaller than band structure calculations [14]. The unfavorable ratio between the small Pauli susceptibility and the other contributions (let alone van Vleck and Landau-Peierls terms) leads to a large uncertainty. The Curie term is attributed to the paramagnetism of Fe ions which are not clustered. Assuming an effective moment $p_{eff}$ = 5.4 to 5.9$\mu_B$, their atomic concentration is estimated to be 60-70 ppm. This paramagnetism causes a Schottky anomaly in the low temperature specific heat.

**Specific heat at high temperature**

Between 15 and 300 K, the specific heat was measured in a calorimeter provided with two concentric adiabatic shields and Pt thermometry [15]. The sample was continuously heated at 15 mK/s and the heat capacity c was obtained from c = P/(dT/dt). Residual losses, typically ±0.1 % of the heater power P, were measured every ≈10 to 20 K and corrected for. The heat capacity of the addenda (sample holder, etc.) was measured separately and subtracted. This calorimeter typically gives 0.01% scatter, 0.2% reproducibility and 0.5% accuracy in fields 0-16 T, using 0.3 gram Cu samples. With $MgB_2$, which has a high Debye temperature, an uncommon problem arises at the lower end of the temperature range: the heat capacity of the 1 mg silicone grease used as an adhesive is as large as 20% of the heat capacity of the 150 mg sample. This is the main limit on the absolute accuracy of the measurement.

Fig. 5 shows the data in the 15-300 K temperature region. Data in B = 0, 10 and 14 T are included. The total scatter is smaller than the thickness of the line. The specific heat reaches only 2/3 of Dulong-Petit equipartition value at room temperature. This indicates a very high effective Debye temperature, $\Theta_{eff} \cong 920$ K at T = 298 K. The inset of Fig. 5 shows the PDOS on a logarithmic scale, compared to a Debye PDOS which has the same moment $<\omega^2>$. The highest phonon frequencies exceed 1000 K, and there is an excess density in the vicinity of 330 K which probably marks the center of gravity of an optical branch.

individual frequencies, their sum, and the $\approx 8000$ data points in the normal state in the form $(C-\gamma T)/T^3$, for two values of the Sommerfeld constant that we consider as upper and lower bounds. The upper bound is the lowest measured value of C/T at the lowest measured temperature in B = 14 or 16 T, i.e. 0.95 mJ/K$^2$gat at 3 K. The lower bound is an extrapolation of the 14/16 T curve to T = 0 using a polynomial expansion, i.e. 0.84 mJ/K$^2$gat.

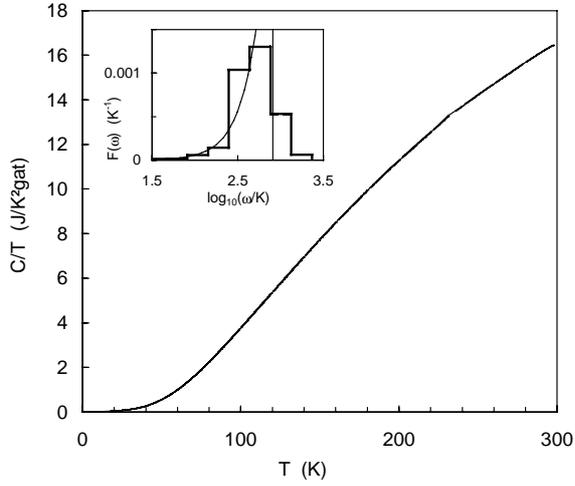

*Figure 5. Specific heat of MgB$_2$ from 15 to 300 K. The >8000 data in B = 0, 10 and 14 T merge into the thickness of the line. Inset: model of the phonon DOS used to fit the data. The weights $w_i$ are divided by the spacing $0.567\omega_i$ between Einstein frequencies $\omega_i$ in the histogram. The curved line is a Debye PDOS having the same moment $<\omega^2>$.*

Figure 6 illustrates how this PDOS was obtained. F($\omega$) was represented by nine $\delta$-functions with frequencies in a geometrical ratio 1.75:1, ranging from 20 K to 1760 K. The weights of these Einstein components were independently fitted to the specific heat curve (Table I). The ratio 1.75:1 was chosen to be large enough to avoid spurious oscillations of the PDOS. The sum of the weights was constrained to 1 (3N modes per gat). Fig 6 shows the contribution of

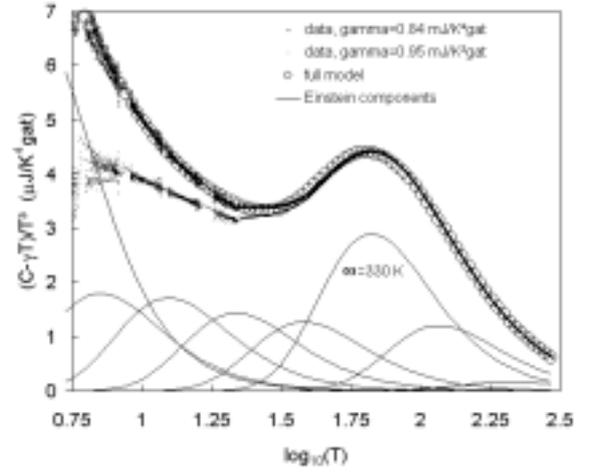

*Figure 6. Plot illustrating the deconvolution of the specific heat into an approximate PDOS. The lattice specific heat is shown in the form $(C_{total} - \gamma T)/T^3$. The lower bound for $\gamma$ is subtracted from the upper data set (diamonds), the upper bound for $\gamma$ is used for the lower data set (crosses). The contributions of the individual Einstein frequencies, equally spaced on the logarithmic temperature axis, are shown in the lower part after fitting their weight (Table I) to the upper set of data. Their total contribution is shown by open circles.*

One notices again the large weight of the mode near 330 K, and an increasing DOS at very low energies which might be due in part to an artifact, i.e. the Schottky contribution of isolated Fe atoms in 14 T, or possibly to a soft phonon mode. For an ideal Debye PDOS, $F(\omega) = 3\omega^2/\omega_D^3$ for $\omega < \omega_D$, $C_{phonon} = \beta_3 T^3$ for $T < \Theta_D/10$, so that the data should align on an horizontal line on the left side of the figure. The departures from this behavior show that an extrapolation of the normal-state specific heat from above $T_c$ into the superconducting state would be uncertain.

*Table I. Position and weight of the Einstein modes used to fit the lattice specific heat of $MgB_2$. The total normal-state specific heat can be reconstructed from the formula C [J/Kgat] = 24.94 $\Sigma_i w_i E(\omega_i/T)$ + 0.00084T. Frequencies are in a geometric progression, $\omega_i = (1.75)^i \omega_0$.*

| i | $\hbar\omega_i/k_B$ (K) | $w_i$ |
|---|---|---|
| 0 | 20 | 0.000103 |
| 1 | 35 | 0.000145 |
| 2 | 61.3 | 0.000739 |
| 3 | 107.2 | 0.00331 |
| 4 | 187.6 | 0.0157 |
| 5 | 328.3 | 0.192 |
| 6 | 574.5 | 0.427 |
| 7 | 1005 | 0.300 |
| 8 | 1759 | 0.0602 |

*Table II. Characteristic parameters of the superconductors $Nb_3Sn$, $YBa_2Cu_3O_7$, and $MgB_2$. Data for $Nb_3Sn$ are compiled from Refs. [2] [16] [17] [18] [19] [20]; $YBa_2Cu_3O_7$ from Refs. [15] [21]; the EDOS for $MgB_2$ is from Ref. [14], tunneling gaps for $MgB_2$ from Refs. [22] [23] [24]; other data are from this work.*
*$M_{gat}$, mass of one gram-atom (one mol of $MgB_2$ is three gat); $V_{gat}$, volume of one gram-atom; $T_c$, average critical temperature obtained by an equal-entropy construction of the idealized specific heat jump; average values of phonon energies $\omega$ as defined in the text; $\gamma$, Sommerfeld constant; N(0), bare band-structure EDOS at the Fermi level; $\lambda_{ep}$, electron-phonon coupling constant; $M\overline{\omega}_2^2$, average lattice force constant; $N(0)<I^2>$, electronic Hopfield parameter; $\Delta C/\gamma T_c$, normalized specific heat jump at $T_c$; $H_c(0)$, thermodynamic critical field at T = 0 obtained by integration of the specific heat difference $C_s-C_n$; $E_c$, measured condensation energy; $0.2364\gamma T_c^2$, BCS value of the condensation energy in the isotropic, weak-coupling limit; $\Delta(0)$, superconducting gap; $(dH_{c2}/dT)_{Tc}$, slope of the upper critical field at $T_c$ as evaluated from the shift of the onset of the specific heat anomaly in fields 1 to 3 T; $H_{c2}(0)$, upper critical field at T = 0; $\xi(0)$, coherence length at T = 0; $\kappa \equiv \lambda/\xi$; $\lambda(0)$, London penetration depth at T = 0; $H_{c1}(0)$, lower critical field at T = 0.*

| | $Nb_3Sn$ | $YBa_2Cu_3O_7$ | $MgB_2$ |
|---|---|---|---|
| $M_{gat}$ [g] | 99.4 | 51.5 | 15.31 |
| $V_{gat}$ [cm$^3$] | 11.1 | 8.0 | 5.83 |
| $T_c$ [K] | 18 | 90 | 36.7 |
| $\omega_{ln}$ [K] | 125 | | 480 |
| $\overline{\omega}_2$ [K] | 174 | | 633 |
| $<\omega^2>^{1/2}$ [K] | 226 | | 808 |
| $\gamma$ [mJ/K$^2$gat] | 13 | ≈1.5 | 0.89 |
| N(0), [eV$^{-1}$atom$^{-1}$spin$^{-1}$] | 0.99 | ≈0.13 | 0.12 |
| $\lambda_{ep}$, from $T_c/\omega_{ln}$ | 1.8 | | (1.07) |
| $\lambda_{ep}$, from $\gamma$/N(0) | 1.8 | ≈1.5 | 0.58 |
| $M\overline{\omega}_2^2$ [eV/Å$^2$] | 5.3 | | 10.9 |
| $N(0)<I^2>$ [eV/Å$^2$] | 9.5 | | 6.3 |
| $\Delta C/\gamma T_c$ | 2.5 | 2? | 0.82 |
| $\Delta C/k_B$, [carriers/cm$^3$] | 3.8×10$^{21}$ | 2.5×10$^{21}$ | 3.3×10$^{20}$ |
| $\mu_0 H_c(0)$ [T] | 0.52 | 1.0 | 0.26 |
| $E_c$ [mJ/cm$^3$] | 108 | ≈400 | 27 |
| $0.2364\gamma T_c^2$, [mJ/cm$^3$] | 90 | ≈360 | 49 |
| $2\Delta(0)/k_B T_c$ | 4.8 | 5 | 1.2-4.2 |
| $-\mu_0(dH_{c2}/dT)_{Tc}$, [T/K] | 1.6 | 2.3 | 0.56 |
| $\mu_0 H_{c2}(0)$ [T] | 25 | 150 | 14 |
| $\xi(0)$ [nm] | 11.5 | 1.5 | 4.9 |
| $\kappa$ | 3.4 | ≈100 | 38 |
| $\lambda(0)$ [nm] | 39 | ≈150 | 185 |
| $\mu_0 H_{c1}(0)$ [T] | 0.13 | ≈0.03 | 0.018 |

The PDOS so obtained is crude, and certainly not unique, but its low order moments are precise and Eliashberg equations do not require any more detailed knowledge. This is because the functional derivative of $T_c$ with respect to variations of the PDOS has a broad maximum near $2\pi T_c$ [5], which is similar to that of the specific heat measured just above $T_c$. Therefore, in some sense, both quantities exhibit the same sensitivity to the PDOS.

The main result of this "inversion" of the specific heat is the determination of average frequencies which are important for superconductivity, as listed in Table II. The logarithmic average $\omega_{ln}$ = 480 K is about 4 times larger than that of the A15 superconductor $Nb_3Sn$ with $T_c$ = 18 K. If the coupling constant $\lambda_{ep}$ of $MgB_2$ was equal to that of $Nb_3Sn$, this high frequency could explain $T_c$ as high as ≈70 K. However, the moment $\overline{\omega}_2$ = 633 K is also 4 times larger, so that the average force constant $M\overline{\omega}_2^2$ increases by a factor of two with respect to $Nb_3Sn$. Now assuming an identical Hopfield electronic parameter η for both compounds, $\lambda_{ep} = \eta/M\overline{\omega}_2^2$ would sink to ≈0.9 in $MgB_2$, bringing $T_c$ back below 30 K. Indeed, the estimate of $\lambda_{ep}$ based on the measured γ and N(0) from band-structure calculations [14] yields only ≈0.6. In other words, phonon frequencies alone cannot readily explain the high $T_c$ of $MgB_2$, provided that the $\alpha^2(\omega)$ function is structureless.

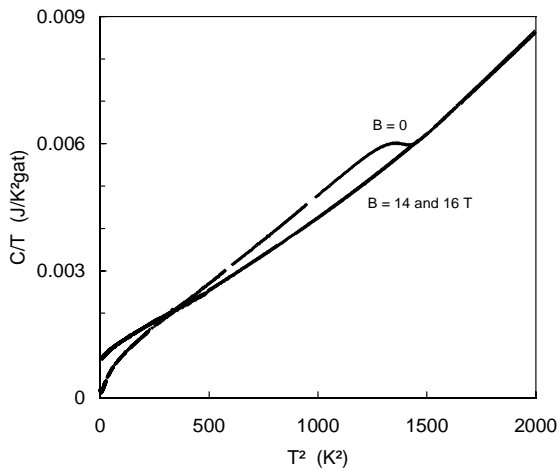

**Figure 7.** Normal-state specific heat of $MgB_2$ in B = 14 and 16 T, and superconducting-state specific heat in B = 0. Note the anomalous curvature of the B = 0 curve at $T \to 0$.

Figure 7 shows both the specific heat in the superconducting state in B = 0 and the specific heat in the normal state in B = 14 and 16 T (the latter data are practically undistinguishable) in the C/T versus $T^2$ presentation. For a classic BCS superconductor with a Debye lattice, the normal state would appear as a straight line $C_n/T = \gamma + \beta_3 T^2$, and the electronic term would vanish exponentially at low temperature in the superconducting state, so that below about $T_c/7$ only the lattice specific heat $C_s/T = \beta_3 T^2$ should survive. Therefore the initial slopes of both curves $C_n/T$ and $C_s/T$ should be equal. This is not the case here. In the superconducting state, the specific heat rather decreases approximately as $C_s/T = A(T^2)^{1/2} + \beta_3 T^2$ (the first term is in fact $C_s = AT^2$, but it appears as a square root function in the C/T versus $T^2$ plot). It is also clear that there is no large residual DOS of normal electrons at T = 0. One also notices a weak negative curvature in the normal state specific heat at $T \to 0$, which corresponds to the excess PDOS at low energies already mentioned, and/or to a Schottky contribution in 14/16 T. From this plot it is clear that the uncertainty in the Sommerfeld constant is small: γ = 0.89 ± 0.05 mJ/$K^2$gat (2.7 mJ/$K^2$/mol, 1530 erg/$K^2 cm^3$). This value lies between those of Cu (0.69 mJ/$K^2$gat) and Al (1.35 mJ/$K^2$gat). With N(0) from Ref. [14], the mass enhancement factor is 1+$\lambda_{ep}$ ≅ 1.6, close to the estimate 1.7 of Ref. [14]. There is no reason to expect an exceptionally large Hopfield parameter η = N(0)<$I^2$> in these conditions. Indeed, if η is estimated as $\lambda_{ep} M\overline{\omega}_2^2$, $\lambda_{ep}$ from the mass renormalization, one finds 2/3 of the value of η for $Nb_3Sn$ (Table II). So neither phonon frequencies, nor the EDOS can account for the high $T_c$ of $MgB_2$.

The specific heat jump at $T_c$ appears rounded in Fig. 7, partly because of the large slope of the lattice background. After removing this background (obtained by averaging the data measured in 10 and 14 T in the same calorimeter), the specific heat jump is better defined (Fig. 8). No doubt that the sharpness of the transition will improve in future samples; however on a relative scale this result is sufficient for thermodynamic studies. The data in fields 0, 0.5, 1, 2 and 3 T were all measured with the adiabatic calorimeter, rather at the low end of the useful range of Pt thermometry (the thermometer used in this calorimeter was calibrated in fields up to 14 T and down to 15 K, using a dielectric thermometer to maintain a constant temperature during field changes).

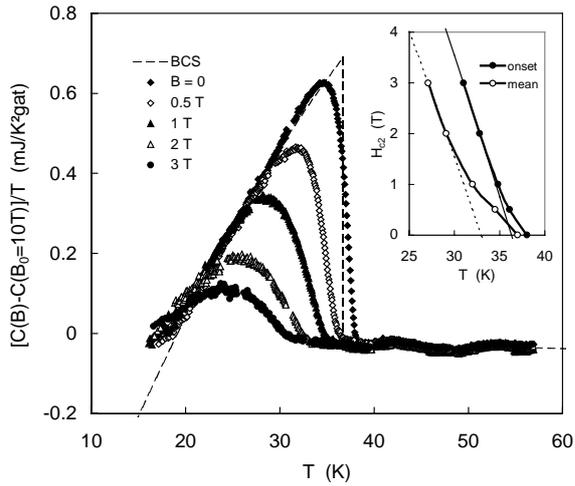

*Figure 8. Specific heat difference between the superconducting and normal state for B = 0, 0.5, 1, 2, and 3 T (adiabatic calorimetry). The normal state is a smoothed average of the B = 10 and 14 T curves. Inset: phase diagram in the H-T plane showing the line of transition onsets (full circles) and the midpoints (open circles).*

The normal- and superconducting- state curves intersect at 19 K, close to $T_c/2$. This crossing is predicted to occur at $0.52\ T_c$ in the isotropic weak-coupling BCS limit [25]. Superconductors with $\lambda_{ep} \approx 1$ instead tend to follow the two-fluid model, $C_s - C_n = 3t^2 - 1$, $t \equiv T/T_c$, with crossing near $0.6\ T_c$ [26]. With a crossing point near $0.8\ T_c$, the cuprate superconductor $YBa_2Cu_3O_7$ differs drastically from classic superconductors, a fact that is not fully understood [27].

The transition is shifted in bulk by the magnetic field to a lower temperature. In a high-$T_c$ superconductor, the transition onset would remain constant, the transition width would increase, and the height of the peak would vanish rapidly with the field. In a classic isotropic type-II superconductor (e.g. $Nb_{77}Zr_{23}$, [26]), the shape of the transition would remain essentially unchanged and the bulk transition would be shifted to lower temperatures. In the present case, the onset of the transition does remain well defined, but the width of the transition increases abnormally. The same observation was made in resistive determinations of the upper critical field, which unlike specific heat may be broadened by flux-flow effects [10] [28]. If the superconductor is not in the clean limit, local variations of the mean free path might broaden the $H_{c2}(T)$ line at high fields. Another more likely explanation is a possible anisotropy of the upper critical field. Grains are randomly oriented in the present polycrystalline sample. If the critical field is a function of the angle θ between $\vec{H}$ and the direction normal to the boron planes, $T_c$ of the grains is distributed between the values defined by the $H_{c2}(T, \theta = 0)$ and $H_{c2}(T, \theta = \pi/2)$ lines. The width of the distribution evidently increases with $H_a$. In this model, the onset of the specific heat jump corresponds to the $H_{c2}(T)$ curve in the "hard" direction, presumably H parallel to the boron planes ($\theta = \pi/2$). We have plotted this onset in the H - T phase diagram (Fig. 8, inset). The slope of this critical field is -0.56 T/K. Assuming a WHH temperature dependence [29], we extrapolate this initial slope to a zero-temperature value close to 14T (the difference between the clean and dirty limit is irrelevant here). This value is consistent with our finding that the specific heat at low temperature is the same in 14 and 16 T. In the same phase diagram, we also show the shift of the midpoint of the calorimetric transition (by midpoint, we mean the center of the straight line extrapolated from the onset up to the B = 0 envelope, following the slope of the specific heat just below the onset). These values of $T_{c2}(H)$ extrapolate to 11.5 T at T = 0 (WHH), and probably represent an angular average of $H_{c2}(T, \theta)$. However, the latter extrapolation is affected by an uncertainty due to the positive curvature at low field. Such a curvature is observed in some classic superconductors, both isotropic (e.g. $Nb_3Sn$ [19]) and anisotropic (e.g. $NbSe_2$ [30]). Fermi surface and pairing anisotropies could produce these effects [31] [32] [33] [34]. Transport measurements of $H_{c2}$ were performed on hot-pressed samples prepared with the same initial powders up to B = 17 T [10]. The midpoints of the resistive transitions follow the midpoints of the specific heat jumps at low fields, with the same positive curvature, but the high field $H_{c2}(T)$ curve is more linear than predicted by WHH, so that $H_{c2}(0)$ = 14 T is also obtained. Similar results are reported by Ref. [28].

The dashed line in the main panel of Fig. 8 shows a BCS fit [25]. Although the BCS curve appears to agree with the data, the Sommerfeld that has to be used in this fit (and which is equal to the jump divided by 1.43) is only 0.5 mJ/K$^2$gat, half the true Sommerfeld constant measured in the normal state. The true dimensionless jump ratio is $\Delta C/\gamma T_c$ = 0.82±10%. $MgB_2$ is *not* a BCS superconductor. Gap anisotropy is known to lower the specific heat jump, essentially because only part of the electrons on the Fermi surface condense at $T_c$. Alternatively, strong coupling would cause a deviation in the *opposite* direction. Some compensation may be present.

On general grounds, one expects a change $k_B$ in the specific heat at $T_c$ for each condensing carrier. According to this criterion, the carrier concentration in $MgB_2$ is found to be one order of magnitude smaller than for $Nb_3Sn$ or $YBa_2Cu_3O_7$ (Table II).

**Specific heat at low temperature**

The specific heat of the *same* sample was measured at low temperature (2-20 K in zero field, 3-16 K in B > 0) in a second calorimeter using a modified relaxation technique as described in the Appendix. Fields B = 0, 1, 3, 14 and 16 T were always applied in the normal state before cooling through the transition in order to avoid metastable states. As already mentioned, the 14 and 16 T data coincide within experimental accuracy, so that $\mu_0 H_{c2}(0) \leq 14$ T. These normal-state data, already included in Fig. 7, determine that $\gamma$ = 0.89 ± 0.05 mJ/K$^2$gat. Previous estimations of $\gamma$ based on extrapolations from high temperatures differ from our result by -60% [35] to +106% [36]. We show later that our results satisfy the entropy condition.

The specific heat difference $(C_s - C_n)/T$ versus T is plotted in Fig. 9. The normal-state data $C_n(T)$ were smoothed before subtraction. The data in 14 and 16 T from the relaxation calorimeter up to 22 K, and the data in 10 and 14 T from the adiabatic calorimeter down to 20 K are also included in this plot after subtracting their smoothed value, in order to show the scatter of the points. The BCS curve fitted at $T_c$ by freely adjusting the parameter $\gamma$ (full line) is clearly inconsistent with the $\gamma$ value found at T → 0. The upturn at low temperature in the B = 0 curve is due to the paramagnetism of Fe impurities, as expected from the Curie component of the susceptibility.

The net area under the $(C_s - C_n)/T$ versus T curve is the entropy difference between the normal and the superconducting state. By integration of the data from 39 K to a variable temperature T < $T_c$, we obtain the curve shown by symbols in the inset of Fig. 9. We have performed no correction for the paramagnetic contribution of impurities, neither in B = 0, nor in B = 14/16 T; the magnetic entropy is present in both fields and partly cancels out. The fact that the entropy difference heads toward zero at T → 0 is important as it asserts the thermodynamic consistency of the data, *including the deviation from the BCS curve below 10 K.*

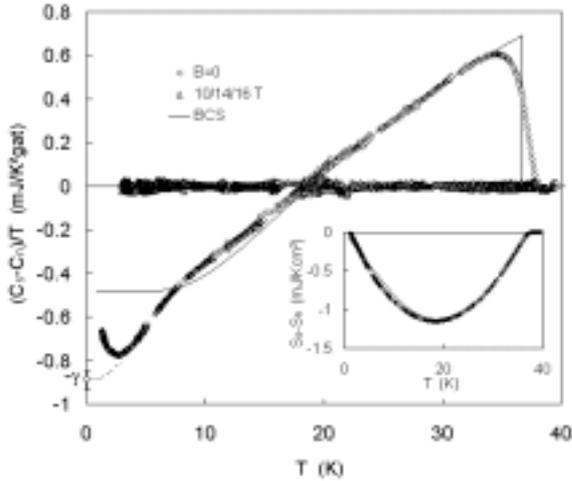

*Figure 9.* Diamonds: specific heat difference between the superconducting state in B = 0 (2 < T < 40 K) and the normal state versus temperature. The normal state is a smoothed curve based on the average of the data in B = 10 T (15 < T < 40 K), 14 T (3 < T < 40 K) and 16 T (3 < T < 16 K). The residual differences between the high field data and this smoothed normal-state curve are shown by triangles. Both adiabatic (T > 16 K) and relaxation (T < 22 K) calorimetry are used. The full line is a tentative BCS fit at $T_c$. Inset: entropy difference obtained by integration of $(C_s-C_n)/T$ from 39 K downwards. The full line is the same BCS fit.

This deviation is one of the central points of the present study. Qualitatively, one may consider that the specific heat difference shown in Fig. 9 results from the superposition of two components. One is a BCS superconductor representing approximately half the volume of the sample. The other component appears to remain in the normal state above $T_c/2$. Then at lower temperature a superconducting gap seems to open progressively in this second component, so that the specific heat difference $(C_s - C_n)/T$ decreases again until it reaches -γ at T = 0 after correction for the magnetic contribution, as it should. Note that there is no second jump, this second gap is distributed.

This is qualitatively the behavior that one expects for a superconductor with an *anisotropic* gap. In Ref. [15], the specific heat was calculated in a simplified model assuming (a) uniaxial anisotropy; (b) an angular dependence of the gap $\Delta(\vec{k}) = \Delta_{max}(1-\cos^2\theta)$, with gap nodes for a motion perpendicular to the B planes; (c) in the low temperature limit, i.e. $\Delta(T) \cong \Delta(0)$. The electronic specific heat in zero field was found to follow the law $C_s/T \propto T$, just as it does in a d-wave superconductor. The underlying reason is that the angle-averaged EDOS in the superconducting state has the same V-shape in both cases; the specific heat is not sensitive to the phase of the order parameter. As seen in Fig. 9, the curve $(C_s - C_n)/T$, which has to point to -γ at T = 0 after correction for the magnetic contribution, shows approximately this behavior (which was also apparent in Fig. 7). In the present case, a finite anisotropy $\Delta(\vec{k}) = \Delta_{max}(1-e\cos^2\theta)$ with e < 1 would probably be more appropriate, but we did not attempt any fit.

The validity of this scenario, which finds some support in the wide spread of values reported for $\Delta(0)$ in tunneling experiments [22] [23] [24], can be further tested by considering the behavior in a magnetic field at low temperature. For a classic isotropic s-wave superconductor (e.g. $Nb_{77}Zr_{23}$ [26]), the specific heat at $T \ll T_c$ is dominated by that of the vortex cores [37] [38], the number of which is proportional to the field. As a result, one observes at $T \to 0$ a field dependence $\lim_{T=0} C_s/T = \gamma(H) = \gamma H/H_{c2}(0)$ after having subtracted the lattice contribution [39].

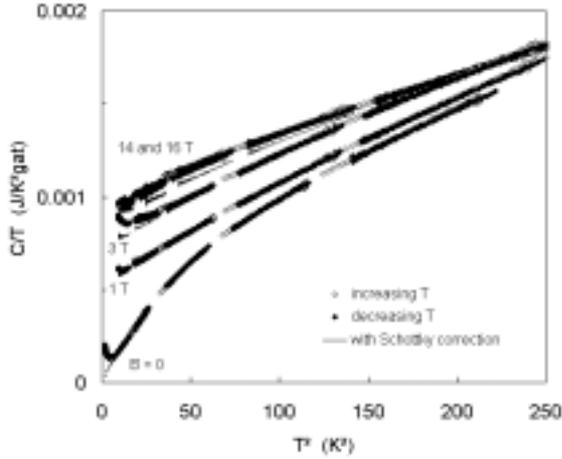

*Figure 10. Total specific heat below 16 K in a C/T versus $T^2$ plot, measured in 0, 1, 3 14 and 16 T fields. The data in 14 and 16 T coincide within experimental scatter. Full symbols are measured with decreasing temperature, open symbols with increasing temperature. The thin lines are corrected for the magnetic Schottky contribution of 120 ppm (atomic) S = ½ spins. Note the non-linearity of the curve in B = 0 and the non-linear field dependence for a given temperature.*

The low temperature specific heat in B = 0, 1, 3, 14 and 16 T is shown in Fig. 10 in the presentation C/T versus $T^2$, so that the lattice contribution essentially adds a positive slope to the intercept $\gamma(H)$ at T = 0. This plot shows two anomalies. The data in B = 0 are not linear in $T^2$, as discussed before. However the essential point is the *non-linear* increase of the specific heat with the field at a given temperature. In $\mu_0 H = 1$ T ($\approx 7\%$ of $H_{c2}$), $\gamma(H)$ already exceeds 50% of $\gamma$. This is again strong evidence in favor of gap anisotropy or different gaps on different sheets of the Fermi surface. The Zeeman and Doppler shifts [40] of the superconducting EDOS in B = 1 T are such that roughly half of the pairs have enough thermal energy at T = 3 K to be excited across the lower parts of the $\vec{k}$-dependent gap.

The contribution of paramagnetic Fe impurities is not believed to be able to explain theses peculiar features in a trivial way. A correction for the contribution of these moments is shown in Fig. 10 by thin lines just below the data. The amplitude of a Schottky anomaly was determined in B = 3 T by using

$$C_{magn}(B,T) = nRx^2 e^x (e^x - 1)^{-2}$$

where $x \equiv 2\mu_e B/k_B T$, $\mu_e$ is the electron magnetic moment and n the atomic concentration of these moments, if C is expressed per gram-atom. We find that with n = 120 ppm, the corrected 3 T curve recovers the expected linear shape at $T \to 0$. Larger values would cause a sharp unphysical downturn and would be definitely inconsistent with the Curie susceptibility measured above $T_c$. Keeping this concentration, we can determine the *effective* interaction field $B_{int}$ that causes the upturn when the *applied* field is zero. The zero field curve, measured down to 2 K and corrected for this magnetic contribution, extrapolates linearly to zero if $B_{int}$ is of the order of 1 T. This procedure does not have to be iterated since the effective field $[(3T)^2+(1T)^2]^{1/2}$ does not differ much from the applied field in 3 T. The 1 T curve, which is measured only above 3 K, is virtually unaffected by this correction. The 14/16 T curves are, and lose some of their negative curvature; however, the raw curve and the corrected curve point almost to the same value of $\gamma$ at T = 0. For simplicity, we did not subtract this magnetic contribution before inverting the specific heat to find phonon frequencies (Fig. 6). The PDOS at the lowest energy that came out of the inversion is possibly in part an artifact due to this contribution, since the normal state curves below $T_c$ are in fact the high field curves. However the corresponding Einstein mode has such a low weight (0.01% of the total number of modes, Table I) that it does not affect the determination of the moments of the PDOS.

**Ginzburg-Landau parameters**

We calculated the thermodynamic critical $H_c(T)$ field according to (cgs units):

$$-\frac{H_c^2(T)}{8\pi} \equiv \Delta F(T) = \Delta U(T) - T\Delta S(T)$$

$$\Delta U(T) = \int_T^{T_c} [C_s(T') - C_n(T')]dT'$$

$$\Delta S(T) = \int_T^{T_c} \frac{C_s(T') - C_n(T')}{T'}dT'$$

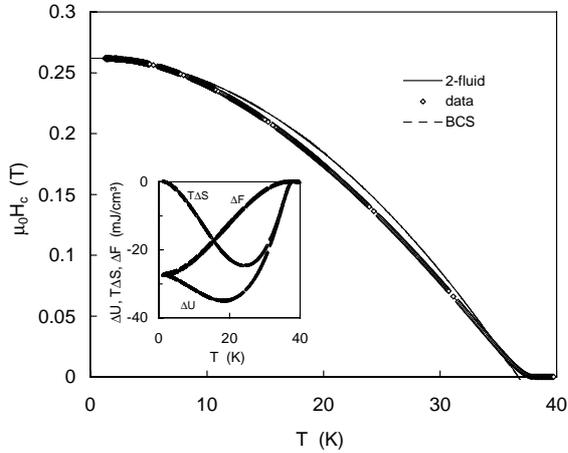

*Figure 11. Diamonds: thermodynamic critical field curve obtained by integration of $C_s-C_n$. Full line: reference two-fluid dependence $h = 1 - t^2$ ($T_c$ is defined by the equal entropy construction shown in Fig. 9). Inset: internal energy difference $\Delta U$, free energy difference $\Delta F$ and entropy difference multiplied by the temperature $T\Delta S$, versus temperature. The condensation energy tends to 27 mJ/cm$^3$ at $T = 0$.*

The result for $H_c(T)$ is given in Fig. 11 and Table II. The inset of Fig. 11 shows separately the differences of internal energy ($\Delta U$), entropy multiplied by the temperature ($T\Delta S$) and free energy ($\Delta F$). The thermodynamic field reaches 0.26 T at $T = 0$, half the value of Nb$_3$Sn, one-quarter of the value of YBa$_2$Cu$_3$O$_7$. Without gap anisotropy, it would be hardly conceivable that the corresponding condensation energy, 27 mJ/cm$^3$, could balance the thermal energy at a critical temperature which is twice that of Nb$_3$Sn and half that of YBa$_2$Cu$_3$O$_7$.

The type-II characteristic parameters listed in Table II are obtained from the Ginzburg-Landau relations (cgs units):

$$H_{c2} = \sqrt{2}\kappa H_c = \frac{\Phi_0}{2\pi\xi^2}$$

$$H_{c1} = \frac{H_c}{\sqrt{2}\kappa}\ln\kappa = \frac{\Phi_0}{4\pi\lambda^2}\ln\kappa$$

where $\Phi_0 = 2.07\times10^{-7}$ G·cm$^2$ is the flux quantum. The second equation is valid for large $\kappa$; here $\kappa = 38$. At $T = 0$, the coherence length is 49 Å and the London penetration depth 1800 Å. The corresponding lower critical field, 180 G, is not apparent in the magnetization curves; the zero-field cooled diamagnetism starts to deviate from $\chi = -1/4\pi$ at the 1% level only between 300 and 500 G (Fig. 2). We verify a posteriori that all specific heat curves in $B > 0$ were measured in the mixed state.

**Conclusion**

Normal- and superconducting-state properties of MgB$_2$ were characterized by thermodynamic measurements (Table II).

We found strong indications suggesting the existence of a multiple-valued gap. These indications are:

- $\lambda_{ep}$ estimated from the *isotropic* Allen-Dynes equation on the one hand, and $\lambda_{ep}$ estimated from the renormalization of the EDOS on the other, differ by a factor of nearly two;

- the normalized curve of the thermodynamic critical field is much closer to the BCS limit than to the strong-coupling curve that is generally found for "high" $T_c$ classic superconductors. In terms of the deviation function, $D(t) \equiv H_c(T)/H_c(0) - (1 - t^2)$, the deviation is *negative* and BCS-like whereas it is usually zero or positive for classic superconductors with $T_c$ in the 10-20 K range, owing to strong coupling corrections;

- the normalized specific heat jump $\Delta C/\gamma T_c \cong 0.8$ is unusually small;

- the condensation energy is low;

- in a magnetic field, the onset of the calorimetric transition for our polycrystal remains sharp whereas the bulk transition is broadened, suggesting an angular dependence of $H_{c2}$ (note that according to Ref. [14] the anisotropy of the Fermi velocity is only 3%);

- the temperature and field dependence of the specific heat at $T \ll T_c$ is strongly reminiscent of that of a d-wave superconductor with line nodes in the gap [40] [41], and essentially different from that of an isotropic s-wave superconductor [37] [38].

Since for a given condensation energy, the superconductor can increase $T_c$ by sacrificing portions of the Fermi surface, we are led to conclude that the high value of $T_c$ in MgB$_2$ is caused both by high average phonon frequencies *and* by a band- or $\vec{k}$-dependence of the gap. Thermodynamic measurements cannot distinguish these two possibilities. Combescot has shown that in the strong-coupling limit, the gap becomes isotropic anyway [42]. This is not in contradiction with the present data, since the mass enhancement factor is only 1.6, well into the weak-coupling regime. These results obtained from a polycrystalline sample can be considered as preliminary and strongly motivate the study of single crystals.

Finally it is instructive to recall the abstract of a prophetic paper by Klein et al.: "(...) it is co1980 ncluded that a metal that does not have a high Debye critical temperature (as do metallic H, B, Be or their alloys) cannot become a high $T_c$ superconductor" [43].

## Acknowledgements

We thank B. Revaz for the development of the relaxation calorimeter, M. Decroux, E. Walker and A. Holmes for communication of unpublished data and help. This work was supported by the Fonds National Suisse de la Recherche Scientifique.

## Appendix: relaxation calorimetry

Our method differs somewhat from conventional procedures, which provide only one value of heat capacity per relaxation. In experiments below 15-20 K, the sample with heat capacity c is connected to a heat sink maintained at a constant temperature $T_0$ by a thermal conductance k in such a way that the time constant of the system, $\tau = c/k$, is of the order of several tens of seconds. A power P is applied to the sample holder for 500 s and then switched off. The sample temperature increases typically up to a final temperature $T_\infty \approx 1.7\ T_0$ and then relaxes to $T_0$. An almost continuous series of heat capacity data is obtained between $T_0$ and $T_\infty$, during heating as well as during cooling. The analysis is based on equation

$$P(T(t)) - P(T(t=\infty))$$
$$= c(T(t))\frac{dT}{dt}\bigg|_t + \int_{T(t=\infty)}^{T(t)} k(T')dT'$$

t = ∞ means in practice t ≈ 10 τ. The derivative is approximated by ($T_{n+1}$ - $T_n$)/($t_{n+1}$ - $t_n$), at a rate of 20 independent points per second. Heat capacity data can be obtained with low error as long as $T(t_{n+1})$ - $T(t_n)$ > 0.001 $T(t_n)$. For stationary states at t < 0 and t >> τ one has

$$P(t=\infty) - P(t<0) = \int_{T(t<0)}^{T(t=\infty)} k(T')dT'$$

The thermal conductance k(T) of the phosphor-bronze thermal link follows approximately k/T = 0.3 μW/$K^2$. A more precise polynomial expansion is determined by a fit, using the above integral equation and 7-9 different temperatures with 3 different powers each. The temperature of the sample is measured by a carbon film calibrated just before the relaxation. This method was thoroughly tested with 9 and 21 mg Ag samples in fields 0-14 T, and found to reproduce standard data to better than 1% [41]. In the low temperature range, the additional heat capacity of the silicone adhesive is relatively less important than at 15-20 K as it decreases with $T^3$; addenda were measured separately and subtracted (in B = 1 T, the addenda/sample heat capacity ratio is 7% at 3 K, 16% at 8 K and 18% at 16 K). An interesting property of this method is the possibility of measuring independently during heating and during cooling. If the heating and cooling curves do not coincide, this means that equilibrium was not reached, for example because the thermal link between the sample and the sapphire platform with the thermometer is not good enough ("$τ_2$-effect"), or because the internal relaxation time of the sample itself is too long compared to the time scale of the measurement (τ). In this unfavorable case, the result for the heat capacity depends on the velocity dT/dt, and data approach equilibrium values only at the end of a relaxation where dT/dt is small. This non-equilibrium situation occurs in the present experiments below 2 K in zero field, and below 3 K in fields larger than 1 T, presumably because of the poor thermal conductivity of the sintered sample. Above these temperatures, data obtained upon heating and cooling coincide on the scale of the figures, and both are included.